\documentclass[11pt,a4paper]{JHEP}
\usepackage{epsfig,bbm}
\usepackage{amssymb}

\def\bX{\overline X}
\def\l{\lambda}

\def\d{\partial}
\def\dw{\partial_w}

\def\Tr{{\rm Tr}}
\newcommand\0{\nonumber}
\newcommand\ee{\end{eqnarray}}	 	%eqnarray
\newcommand\be{\begin{eqnarray}}
\newcommand\ba{\begin{array}}			%array
\newcommand\ea{\end{array}}

\newcommand\e{{\rm e}}

\preprint{SISSA/51/99/EP/FM\\hep-th/9905092}

\title{Heterotic Matrix String Theory and Riemann Surfaces}

\author{G. Bonelli, L. Bonora, F. Nesti, A. Tomasiello\\
	International School for Advanced Studies (SISSA/ISAS)\\
	Via Beirut 2--4, 34014 Trieste, Italy, and INFN, Sezione di Trieste\\
	E-mail: \email{bonelli@sissa.it}, \email{bonora@sissa.it}, 
	\email{nesti@sissa.it}, \email{tomasiel@sissa.it}}

\abstract{We extend the results found for Matrix String Theory to
Heterotic Matrix String Theory, i.e. to a 2d $O(N)$ SYM theory with
chiral (anomaly free) matter and ${\cal N}=(8,0)$ supersymmetry. We
write down the instanton equations for this theory and solve them
explicitly. The solutions are characterized by branched coverings of
the basis cylinder, i.e. by compact Riemann surfaces with
punctures. We show that in the strong coupling limit the action
becomes the heterotic string action plus a free Maxwell
action. Moreover the amplitude based on a Riemann surface with $p$
punctures and $h$ handles is proportional to $g^{2-2h-p}$, as expected
for the heterotic string interaction theory with string coupling
$g_s=1/g$.}

\begin{document}

\section{Introduction}\label{1}

The purpose of this paper is to extend to Heterotic Matrix String
Theory (HMST),\linebreak \cite{DF,KS,lowe,BM,rey,horava,kabat,govin,krogh,BGS}
the results obtained recently for Matrix String Theory (MST),
\cite{wynter,GHV,bbn1,bbn2,bbnt} (see also the related papers
\cite{winter2,BCDP,kostov,makee,gri,sugino}). In the latter it was
shown what had been previously conjectured in~\cite{motl,BS,DVV,HV},
namely that MST, i.e. $U(N)$ SYM theory in two dimensions with adjoint
matter and ${\cal N}=(8,8)$ supersymmetry, describes in the strong
coupling limit IIA superstring theory and all its string
interactions. The relevant Riemann surfaces are dynamically generated
via BPS instantons, which are referred to throughout as {\it stringy
instantons}. Each Riemann surface (with $p$ punctures and $h$ handles)
is associated to a corresponding string process in an obvious way.  To
verify the correctness of this association the action was expanded
about an instanton; it was shown in~\cite{bbn2} that the resulting
amplitude is proportional to $g_s^{-\chi}$, where $g_s$ is the string
coupling constant (i.e. the inverse of the Yang-Mills coupling) and
$\chi=2-2h-p$ is the Euler characteristic of the Riemann surface.
This is what one expects from perturbative string interaction
theory. Relying on this basic result, in~\cite{bbnt}, we set out to
analyze the moduli space of stringy instantons and found that it does
not exactly coincide with the moduli space of compact Riemann surfaces
with punctures, which is relevant to closed string theory, but it
mimics it very closely: in fact it is a discretized version of the
latter in the sense that some of the moduli are discrete. However
these moduli become continuous in the $N\to\infty$ limit.  We argued
in~\cite{bbnt} that in this limit the moduli space of stringy
instantons does reproduce the moduli space of Riemann surfaces
relevant to the type IIA theory.

In the present paper we extend the above result to the HMST. The
latter is an $O(N)$ SYM theory with chiral (anomaly free) matter and
${\cal N}=(8,0)$ supersymmetry. We first write down the instanton
equations for this theory. Then we show how one can solve them
explicitly. The solutions are of course different from the MST case,
but they span the same kind of branched coverings of the basis
cylinder, i.e. the same kind of compact Riemann surfaces with
punctures, as in MST.  We then expand the action about any such
instanton in the strong coupling limit. The details are different from
the MST case and somewhat subtler (see Appendix), however the final
result is similar to MST.  The strong coupling limit action is the
heterotic string action plus a free Maxwell action. Moreover the
amplitude based on a Riemann surface with $p$ punctures and $h$
handles is proportional to $g_s^{2h+p-2}$, as expected for the
heterotic string interaction theory. At this point the analysis of the
moduli space of stringy instantons goes through as
in~\cite{bbnt}. This moduli space is a discretized approximation of
the moduli space of Riemann surfaces in heterotic string theory. The
approximation gets better and better as $N$ becomes larger.

The paper is organized as follows. Section~\ref{sec:2} is devoted to
heterotic stringy instantons. In section~\ref{sec:3} we expand the
action around any instanton solution and compute the strong coupling
limit. The Appendix contains a few details of the relevant
calculations.

\section{The 2d SYM model and classical supersymmetric configurations}
\label{sec:2}

The Heterotic Matrix String Theory (HMST) is specified by an $O(N)$
SYM model in a 2d Euclidean space-time. We will choose $N$ even,
$N=2n$. The action is
\begin{eqnarray} 
S&=&\frac{1}{\pi} \int d^2w \,\Big\{\Tr \Big(
D_w X^i D_{\bar w} X^i + \frac{1}{4g^2} F_{w\bar w}^2 -
\frac{g^2}{2}[X^i,X^j]^2+i (\theta_s D_{\bar w} \theta_s - \theta_c
D_w \theta_c)-  \0\\
&&~~~~~~~~~~~~~~~~~~~{} 
-2g \theta_s \Gamma_i [X^i,\theta_c] \Big)+ i \chi D_w \chi \Big\}\,,\label{eSYM}
\end{eqnarray}
where we use the complex notation
\begin{eqnarray}
w= \frac {1}{2} (\tau +i \sigma)\,,\qquad 
\bar w = \frac {1}{2} (\tau - i \sigma)\,,
\qquad A_w= A_0-iA_1\, ,\qquad A_{\bar w}= A_0+iA_1\,,\0
\end{eqnarray}
where $\sigma$ and $\tau$ are the world-sheet coordinate on the
cylinder.  In~(\ref{eSYM}) $g$ is the gauge coupling, $X^i$ with
$i=1,\ldots,8$ are real symmetric $N\times N$ matrices.  $A_0 , A_1$
are real antisymmetric $N\times N$ matrices. The covariant derivative
is defined as $D_w X^i = \partial_w X^i - [A_w, X^i]$. $F_{w,\bar w}$
is the gauge curvature.  Summation over the $i,j$ indices is
understood.  $\theta_s(\theta_c)$ represents 8 $N\times N$ symmetric
(antisymmetric) matrices whose entries are 2D real spinors and
simultaneously transform in the ${\bf 8_s}$ and ${\bf 8_c}$
representations of $SO(8)$. $\chi$ are 2D fermions transforming
according to the fundamental representation of $O(N)$ and the
fundamental representation of $SO(32)$. The matrices $\Gamma_i$ are
the $16\times 16$ $SO(8)$ gamma matrices. For definiteness we will
write the matrices $\Gamma_i$ in the form
\begin{equation}
\Gamma_i = \left(\matrix {0 & \gamma_i\cr
                 \tilde \gamma_i & 0\cr} \right),\0
\end{equation}
and $\gamma_i,\tilde\gamma_i$ are the same as in Appendix 5B
of~\cite{GSW}.  All the fields in HMST are {\it periodic} on the
cylinder.

The action~(\ref{eSYM}) is chiral, however the fermion representations
are such that the gauge anomaly vanishes (see also Appendix). It is
invariant under the supersymmetry transformations (${\cal N}= (8,0)$
supersymmetry)
\begin{eqnarray}
\delta X^i &=& \frac{i}{g} \epsilon_c \tilde\gamma^i \theta_s\0\\
\delta \theta_s &=& -\frac{1}{g}D_w X^i
\gamma_i\epsilon_c\0\\
\delta \theta_c &=& (\frac{i}{2g^2} F_{w\bar w} -
\frac {i}{2} [X^i,X^j]\tilde\gamma_{ij}) \epsilon_c \0\\[.5ex]
 \delta A_{\bar w}&=&-2\epsilon_c \theta_c\0\\
 \delta A_w &=& 0\,, \qquad\qquad \delta \chi=0\,,\label{susy}
\end{eqnarray}              
where
\begin{eqnarray}
\gamma_{ij} = \frac {1}{2} (\gamma_i\tilde\gamma_j - \gamma_j\tilde \gamma_i)\,,
\qquad \gamma_{ij} = \frac {1}{2} (\tilde\gamma_i\gamma_j - 
\tilde\gamma_j\gamma_i)\,.\0
\end{eqnarray}
 
The HMST, in the strong coupling limit, is expected to represent
heterotic string theories. We can enrich the content of (\ref{eSYM})
by introducing Wilson lines.  The corresponding term in the integrand
of~(\ref{eSYM}) is $\chi B\chi= \sum_{a=1}^{32} \chi^a B_{ab}\chi^b$,
where $B$ is a real antisymmetric matrix.

A specific choice of $B$ leads to a theory in which, for example, $SO(32)$ is
broken to $SO(16)\times SO(16)$. Via a T-duality transformation this
theory can be related to the {\it ten-dimensional} $E_8\times E_8$ heterotic 
theory broken down to $SO(16)\times SO(16)$ by another suitable Wilson line,
see~\cite{kabat,govin,krogh,BGS}. In the absence of Wilson lines it is
expected to represent $SO(32)$ heterotic string theory compactified to
nine dimension on a very small circle. 

All this, as well as the s-duality connection with type IA and type IB
theories~\cite{PW,HW}, is well-known. However the introduction of
Wilson lines does not affect our subsequent derivation. Therefore we
will drop them throughout and resume them at the end of our
derivation.

\subsection{Interpretation of the strong coupling states}

The naive strong coupling limit ($g\to \infty$) in the action, after
rescaling $A\to gA$, tells us that all the $X^i$, $A_w,A_{\bar w}$ and
$\theta$ commute, and $\chi A_w\chi=0$.
  
Denoting with a bar the fields in the strong coupling limit, one sees
that there are two types of solutions. Let us denote by ${\bf 1}$ the
$2\times 2$ identity matrix and by ${\bf \epsilon} = \left( \matrix{0
&1\cr -1&0\cr}\right)$. The first type of solution is: $\bar A_w=\bar
a\otimes \epsilon,\bar\theta_c= \bar\vartheta_c\otimes \epsilon$,
$\bar X^i=\bar x^i\otimes {\bf 1}$ and $\bar\theta_s=\bar
\vartheta_s\otimes {\bf 1}$, where $\bar a, \bar x^i, \bar
\vartheta_c$ and $\bar \vartheta_s$ are diagonal matrices. Moreover
$\bar\chi$ is such that $\bar\chi \bar A_w\bar\chi=0$, which implies that
half of the degrees of freedom of $\bar\chi$ must vanish. For a reason
that will be explained below, we actually scale out $\bar\theta_c$ by
multiplying it by a suitable power of $1/g$, so that in the above
formulas it is understood $\bar \vartheta_c=0$.

A second group of solution is characterized by $\bar
A_w=0$. Consequently $\bar X^i$ and $\bar\theta_s$ are generical
diagonal matrices (without the two by two identification of
eigenvalues as before). Supersymmetry then requires that $\bar
\theta_c=0$.

In the following we refer to these two group of solutions as first and
second branch, respectively. Although one can hardly attach too much
importance to these naive elaborations, they actually turn out to be
very suggestive and not at odds with the results of the more
appropriate treatment presented later.

It is well-known that in the strong coupling theory there is a
residual gauge freedom, which contains the Weyl group of $O(N)$ and
allows a variety of boundary conditions. Each one of them, in the
first branch, defines a string configurations.  Let us consider, for
example, a solution from the first branch. We have in particular $\bar
X^i= Diag(x_1^i,\ldots,x_n^i)\otimes {\bf 1}$.  The distinct
eigenvalues of the latter can be interpreted as free strings of
various lengths. For instance,\footnote{Here we use the same explicit
example as in~\cite{bbn1} in order to point out the analogies and
differences between the two cases} let us consider the effect on $\bar
X^i$ of the element $\hat{\cal P}\equiv {\cal P}\otimes {\bf 1}$ of
the Weyl group of $O(N)$, where
\be {\cal P}= \left(\matrix{0& 0& \ldots & \ldots & 1\cr 
	1 & 0 & \ldots & \ldots & 0 \cr
	0 & 1 & 0 & \ldots & 0 \cr
	\ldots & \ldots & \ldots& \ldots &\ldots\cr 
	0 & 0 & \ldots & 1 & 0 \cr}\right).\0 
\ee 
The boundary condition $\bar X^i(2\pi) = \hat{\cal P} \bar X^i(0)
\hat{\cal P}^{-1}$ implies that $x^i_k(2\pi) =x^i_{k-1}(0)$, and so
the $x^i_k$ form a unique long string of length $2\pi n$.  A parallel
interpretation holds for the fermionic degrees of freedom.  We see
that, apart from the gauge field, we get the spectrum of the free
heterotic superstring. The gauge degrees of freedom need further
specification. Here we only anticipate that the gauge field will turn
out to be essential for the interpretation of the strong coupling
limit of HMST as heterotic string {\it interaction} theory.

One may wonder at this point why above we scaled out $\bar\theta_c$
and not in its stead the full set of $\bar\chi$ degrees of freedom. We
would have obtained in this way, apart from the gauge degrees of
freedom, the spectrum of the free type IIA superstring theory. The
answer is that there are various indications that the strong coupling
limit in this case would be discontinuous. For example, the HMST is a
chiral fermionic theory, while the strong coupling limit would not be
chiral. Moreover the supersymmetry transformations inherited by those
of HMST (see~(\ref{susy}) are not those of the type IIA. Therefore we
exclude the possibility of such a strong coupling limit.

We can proceed similarly also for the second branch solutions. They
have apparently a spectrum which coincides with the heterotic
superstring spectrum. However, one sees immediately that the residual
discrete gauge transformations, which form the Weyl group of $O(N)$,
are not the expected ones for a heterotic string interpretation of
this branch. This is a spy of of the fact that, as will become clear
later on, the strong coupling limit of the second branch, if it
exists, cannot be interpreted as a string theory.

\subsection{2D instantons and Hitchin systems}

We look now for classical Euclidean supersymmetric configurations that
preserve half supersymmetry. To this end we set $\theta=\chi=0$ and
look for solutions of the equations $\delta \theta=0$, i.e.,
from~(\ref{susy}),
\begin{eqnarray}
\left(\frac{i}{2g^2} F_{w\bar w} -
\frac {i}{2} [X^i,X^j]\tilde\gamma_{ij}\right) \epsilon_c =0\,,\quad 
D_w X^i \gamma_i\epsilon_c=0\,.\label{insteq0}
\end{eqnarray}
Solutions of these equations that preserve one half supersymmetry are
the following ones. Set $X^i =0$ for all $i$ except two, for
definiteness $X^i\neq 0$ for $i=1,2$; remark that $\gamma_{12}$ is an
antisymmetric $8\times 8$ matrix, and $\gamma_{12}^2=-1$ and therefore
its eigenvalues are $\pm i$ ( moreover $\tilde
\gamma_{12}=\gamma_{12}$). It is easy to show that there exists
$\epsilon$, with four independent components, such that
\begin{eqnarray}
\gamma_{12} \epsilon = i\epsilon, \qquad
\gamma_1 \epsilon= -i\gamma_2 \epsilon.\0
\end{eqnarray}
We replace this $\epsilon$ in eq.~(\ref{insteq0}).
It is convenient to introduce the complex notation $X=X^1+iX^2$, $~\bar X=
X^1-iX^2= X^\dagger$. $X, \bar X$ are complex symmetric matrices. 
Then the conditions to be satisfied in order to preserve one half supersymmetry 
are
\begin{eqnarray}
&&F_{w\bar w} +  g^2 [X, \bar X] =0, \quad\quad D_w X=0, 
\label{insteq}
\end{eqnarray}
It is easy to verify that, if non-trivial solutions to such equations exist,
they satisfy the equations of motion of the action~(\ref{eSYM}). The 
action with $\theta=0,\chi=0, X^i=0$ for $i=3,\ldots8$ can be normalized
as follows
\begin{eqnarray}
S_{inst}=\frac{1}{2\pi} \int d^2w \,\Tr \left(-
XD_w D_{\bar w}\bar X - \bar XD_w D_{\bar w} X+ \frac{1}{2g^2} F_{w\bar w}^2 +
\frac{g^2}{2}[X,\bar X]^2 \right).\label{instaction}
\end{eqnarray}
It is elementary to prove that $S_{inst}$ vanishes in correspondence to 
solutions of~(\ref{insteq}).

\section{Heterotic stringy instantons and Riemann surfaces}
\label{sec:3}

\subsection{Heterotic stringy instantons}

We recall that $N$ is even, $N=2n$. We want to construct solutions of
eqs.~(\ref{insteq}) that interpolate between any two asymptotic string states 
($w=\pm \infty$) as the ones considered in sec.2.1.
It turns out that the right (heterotic stringy) strong coupling solutions 
of~(\ref{insteq}) are of the type:
\be
X \to X^{(b)},\quad\quad A\to  A^{(b)}=0\,, 
\label{stringy}
\ee
where $X^{(b)} = \hat X\otimes {\bf 1}$
and $\hat X= {\rm Diag}(x_1,\ldots, x_{n})$.  
We construct the instanton that reduces to it in the strong coupling limit
in close analogy with~\cite{bbn1,bbn2}. 
 
Our purpose is to construct a symmetric matrix $X$ and an antisymmetric 
connection $A$ that satisfy~(\ref{insteq}).
Let us start from the canonical $n\times n$ matrix $M$ whose eigenvalues 
coincide with those of $\hat x$,
\be
M=\left(\matrix{-a_{n-1}& -a_{n-2}&\ldots & \ldots & -a_0\cr 
		1 & 0 & \ldots & \ldots & 0 \cr 
		0 & 1 & 0 & \ldots & 0 \cr 
		\ldots & \ldots & \ldots& \ldots & \ldots \cr 
		0 & 0 & \ldots & 1 & 0 \cr
}\right).\label{can}
\ee
We map the problem from the cylinder to the disk: $w\to \bar z= e^w$,
and require $\d_{\bar z} M=0$, and therefore $\d_{\bar z} a_i=0$ for
each $i=0,\ldots,n-1$, after mapping .  The matrix $M$ defines a
branched covering of the cylinder ${\cal C}$, which is a Riemann
surface with punctures, as discussed at length
in~\cite{bbn1,bbn2,bbnt}.

Now we identify the eigenvalues of $\hat X$ with those of $M$.  We
diagonalize $M$
\be
M= S \hat M S^{-1}
,\quad\quad \hat M = {\rm Diag}(x_1,\ldots, 
x_{n})\0
\ee
by means,~\cite{wynter}, of the following matrix $S\in SL(n,{\mathbb C})$:
\be
S= \Delta^{-{1\over n}} 
		\left(\matrix{x_1^{n-1}&x_2^{n-1}& \ldots & \ldots &x_n^{n-1}\cr
		x_1^{n-2}&x_2^{n-2}& \ldots & \ldots &x_n^{n-2}\cr
		\ldots & \ldots & \ldots& \ldots & \ldots \cr
                1 & 1 & \ldots & \ldots & 1 \cr
}\right),\label{S}
\ee
where 
\be
\Delta = \prod_{1\leq i<j\leq n}(x_i - x_j)\,.\label{Delta}
\ee

Next we set
\be 
S= B Q\,,\0
\ee
where $B = \sqrt{S \tilde S} = \tilde B$. Here tilde stands for
transposition.  One has $Q\tilde Q = B^{-1} S \tilde S B^{-1}=1$. In
other words we decompose $S$ into the product of a complex symmetric
matrix $B$ and a complex orthogonal matrix $Q$.

As a consequence 
\be
M_s = B^{-1} M B = Q \hat M Q^{-1}\label{sym}
\ee
is a symmetric complex matrix. We choose $M_s$ as our starting point
in the construction of the instanton, in the same way as we used $M$
in~\cite{bbn1,bbn2}.  At this point it would seem that we have to
accompany $M_s$ with a corresponding connection $A_a = - Q \d
Q^{-1}$. However one can prove that the latter identically
vanishes. This follows from the fact that $Q$ and $Q^{-1}$ contain
strictly fractional singularities in $z$, say $(z-a)^{k/j}$ with $k,j$
relatively prime integers, but never $(z-a)^{-l}$, with $l$ positive
integer.  Therefore, in the sense of complex distributions, $\d_{\bar z}
Q^{-1}=0$.

Now the complex orthogonal matrix $Q$ can be decomposed as follows
\be
Q = H O, \quad\quad H = \sqrt {Q Q^\dagger}\,.\label{O}
\ee
As a consequence, $H=H^\dagger$, $OO^\dagger =1$ and $O\tilde O =1$.
Therefore $O\in O(n,{\mathbb R})$. 

Our complete ansatz for the instanton is defined by the couple
$X=Y^{-1} M_sY \otimes {\bf 1}$ and $A= -Y^{-1} \d_w Y\otimes {\bf
1}$, where $Y= H L$, and $L$ is a complex orthogonal matrix. In
particular
\be
X = Y^{-1} M_sY\otimes {\bf 1}= L^{-1}O \hat M O^{-1} L\otimes {\bf 1}\,.
\label{X}
\ee
Since $L$ is a complex orthogonal matrix, $X$ is symmetric and $A$ is
antisymmetric, as desired.

\pagebreak[3]

The points where $\Delta$ vanishes, i.e. where two eigenvalues
coincide, are branch points of the covering defined by the matrix
$M$. They are therefore characterized by a monodromy. Let us check
that the solutions we propose are not affected by such monodromies,
i.e. are single-valued.  Notice that under a monodromy transformation
around any point of the basis cylinder, we have
\be
S \to S\Lambda\,,\qquad B\to B\,,\qquad Q \to Q\Lambda\,, \qquad H\to H\,,\qquad O\to
O\Lambda\,, \qquad \hat M \to \Lambda^{-1} \hat M \Lambda\0\,,
\ee
where $\Lambda$ is the monodromy matrix about the given point (one
such monodromy matrix is, for example, the matrix ${\cal P}$
introduced above).
 
Therefore $M_s$ and $X$, as well as $A$, are single-valued provided
$L$ is. Let us see how one can construct such an $L$.

In order for our ansatz to satisfy~(\ref{insteq}), the entries of the
complex orthogonal matrix $L$ must be single-valued fields that
satisfy field equations of the WZNW type, with delta-function-type
sources at the branch points. We have already discussed the existence
of solutions of such equations in~\cite{bbn1,bbn2,bbnt}.  We have seen
that in the strong coupling limit $L$ tends to 1 outside an arbitrary
small neighborhood of the branch points.  Therefore, excluding a small
neighborhood of the branch points, we can set
\be
X= O \hat X O^{-1}\otimes {\bf 1}\,, \quad\quad A = - H^{-1}\dw H \otimes {\bf 1}
= - O\dw O^{-1}\otimes {\bf 1}\0
\ee
since, as we have seen, $\dw Q=0$. Therefore in the strong coupling
limit, ouside the branch points, we recover $A^{(b)}$ and $X^{(b)}$ up
to a gauge transformation.\footnote{Of course, if $O\in O(n)$,
$O\otimes {\bf 1}\in O(N)$.}

We also remark that eqs.~(\ref{insteq}) are invariant under orthogonal
$O(N)$ transformations. Therefore the instantons solutions we have
found are understood to be defined up to arbitrary orthogonal $O(N)$
transformations.

\subsection{Other branches}

With the term strong coupling solutions we mean the solutions
of~(\ref{insteq}) in which $F_{w\bar w}=0$ and $[X,\bar X]=0$
separately. There are many other possible strong coupling solutions of
(\ref{insteq}) beside~(\ref{stringy}). Therefore it is important to
understand why in this paper we limit ourselves to~(\ref{stringy}).
The strong coupling configurations split into two branches, just as
the solutions studied in naive strong coupling limit in sec.2.1.  The
first branch is given, in the diagonal representation, by
\be 
A_{ w} =A_w^{(b)}= \hat A \otimes {\bf \epsilon}\,,\qquad X\equiv X^{(b)} = 
\hat X\otimes {\bf 1}\,,
\label{branch2}
\ee
where $\hat A$ and $\hat X$ are diagonal matrices. The second branch
corresponds to configurations $(X^{(b)}, A^{(b)})$ in which
$A^{(b)}=0$ and $X^{(b)}$ is a generic diagonal matrix (i.e. without
two by two identification of the eigenvalues as
in~(\ref{stringy}). This classification closely parallels the one
found in section 2.1 in the naive strong coupling limit.
 
The strong coupling solutions~(\ref{stringy}) correspond to the
intersection of the two branches. In the same way as above we
constructed the instantons corresponding to any strong coupling
solution~(\ref{stringy}), we can construct instantons corresponding to
any strong coupling solution in the two branches. This has been done
and will be reported elsewhere. However a generic instanton in the two
branches needs not lead in the strong coupling limit to the heterotic
string theory. The instantons in the intersection (i.e.  the stringy
instantons) do.

Let us clarify what we mean by this.  As is clear
from~\cite{bbn1,bbn2} and from the next section, our aim is to expand
the action about a given instanton solution and to extract the strong
coupling limit. The instanton is really {\it stringy} if the Riemann
surface it contains (as a branched covering of the cylinder) is in
accord with the string interpretation as a scattering process, in the
strong coupling limit. This requires that the corresponding amplitude
be proportional to $g_s^{-\chi}$, where $\chi$ is the Euler
characteristic of the Riemann surface, as explained in the
introduction. This factor can only come from a Maxwell field theory on
the covering of the cylinder, which, in turn, can only be a
consequence of a surviving part of the original gauge symmetry of the
theory.  We recall that in~\cite{bbn1,bbn2} the $U(N)$ gauge symmetry
of the theory breaks down, in the strong coupling configurations on
the basis cylinder, to $(U(1))^N$, and that this is the basis for the
persistence of a $U(1)$ gauge symmetry on the covering in the strong
coupling limit.  In the instantons constructed in the previous
subsection, the strong coupling configurations~(\ref{stringy}) break
down the gauge symmetry $O(N)$ in such a way that a $(O(2))^n$
symmetry survives. This, as will be seen, guarantees that a Maxwell
theory will survive on the corresponding covering in the strong
coupling limit.  This is the reason why we limit ourselves in this
paper to strong coupling configurations~(\ref{stringy}) represented by
$ X^{(b)}$'s in which the eigenvalues are identified two by two,
i.e. are in the intersection of the two branches.

Should we consider instead diagonal $X^{(b)}$'s in which all
eigenvalues are different (generic solutions of the second branch), any
non-discrete gauge symmetry would be destroyed in the strong coupling
limit. This would not leave any room for a Maxwell field on the
covering, therefore such configurations cannot trigger a string
interpretation.  As for generic solutions of the first branch, they
contain a non-vanishing background gauge potential. The corresponding
coverings may possibly be interpreted in terms of scattering of
heterotic strings with non-perturbative objects.

In conclusion, only in the intersection of the two branches a genuine
heterotic string interpretation of the strong coupling limit seems to
be possible. That is the reason why we called heterotic stringy the
instantons corresponding to the intersection between the two branches.

\subsection{Lifting to the spectral covering}

Following~\cite{bbn2}, we show now that the HMST in the strong YM
coupling limit, can be lifted to the covering $\Sigma$ of the basis
cylinder.  To this end we first rewrite the action in the following
form
\be
S&=&\frac{1}{\pi} \int d^2w \,\left\{\Tr \left(
D_w X^I D_{\bar w} X^I 
-\frac{g^2}{2}[X^I,X^J]^2
-g^2[X^I,X][X^I,\bX]
+D_w X D_{\bar w} \bX 
\right.\right.\0\\
&& \left.\left.
+\frac{1}{4g^2}\left( F_{w\bar w} + g^2[X,\bX]\right)^2
 +i (\theta_s D_{\bar w} \theta_s - \theta^+_c
D_w \theta^+_c) -g \theta^T \Gamma_i [X^i,\theta] \right)
+i\chi D_w\chi\right\}\0\,,
\ee
where $I= 3,4,...,8$.  We now expand the action around a generic
instanton configuration writing any field $\Phi$ as
\be
\Phi=\Phi^{(b)}+\phi^{\mathfrak t}+\phi^{\mathfrak n}\equiv 
\Phi^{(b)}+\phi\equiv \Phi^\circ +\phi^{\mathfrak n}\,,
\label{phi}
\ee
where $\Phi^{(b)}$ is the background value of the field at infinite
coupling, $\phi^{\mathfrak t}$ are the fluctuations along the Cartan
directions (for $A$ and $\theta_c$) or the directions which commute
with the Cartan generators (for $X^i$ and $\theta_s$), while
$\phi^{\mathfrak n}$ are the fluctuations along the complementary
directions in Lie algebra ${\mathfrak o}(N)$ and in the relevant
representations. So, in particular,
\be
a_w^{\mathfrak t}= a_w^{\mathfrak d}\otimes \epsilon,
\qquad x^{i{\mathfrak t}}= x^{i{\mathfrak d}} \otimes {\bf 1}\,,
\qquad \theta^{\mathfrak t}_c = \vartheta_c^{\mathfrak d} \otimes {\epsilon}\,,
\qquad \theta_s^{\mathfrak t}= \vartheta_s^{\mathfrak d} \otimes {\bf 1}\,,
\label{diagonal}
\ee 
where $a^{\mathfrak d}_w,x^{i{\mathfrak d}}, \vartheta_s^{\mathfrak
d}, \vartheta_c^{\mathfrak d}$ are diagonal $n\times n$
matrices. Therefore for instance $a_w^{\mathfrak n}$ has $2n(n-1)$
components, while $x^{i{\mathfrak n}}$ have $2n^2$ components each.

As for $\chi$ the above splitting has a specific meaning. As a vector
in the fundamental representation of $O(N)$, we split it as follows
\be
\chi^{\mathfrak t} = 
\frac{\chi_0}{\sqrt 2}\otimes \left(\matrix {1\cr 1\cr}\right), \qquad
\chi^{\mathfrak n}= 
\frac{\chi_1}{\sqrt 2} \left(\matrix{1\cr -1\cr}\right),
\label{chi}
\ee 
where $\chi_0,\chi_1$ are $n$-component vectors. This splitting is
actually rather arbitrary: we could replace $\chi_0$ and $\chi_1$ with
any two linearly independent combinations of them. However the final
result below would not change.

In the strong coupling we consider the above action on the base space
${\cal C}_0$, i.e. the initial cylinder with a small disk cut out
around any branch point (in other words we insert a regulator defined
by the radius of such disks). Then the only field with a
non-vanishing background part is $X^{(b)}= \hat X \otimes {\bf 1}$
and its complex conjugate, since $A^{(b)}=0$ dressed with the
orthogonal gauge transformation $O$, see~(\ref{O}).  We can get rid of
$O$ by means of a gauge transformation in the action. But we have to
exercise some care because in so doing we would $O$-rotate all the
fluctuations (both Cartan and non-Cartan). In order to simplify
things we will understand that the fluctuations in~(\ref{phi}) are
defined up to the gauge transformation $O^{-1}$.

Now, since the value of the action on the background is zero and since
the background is a solution of the equation of motion, the expansion
of the action starts with quadratic terms in the fluctuations. Next we
proceed, exactly, as in~\cite{bbn2}.  We use the same gauge fixing
\be
{\cal G}=D^\circ_w a_{\bar w} +D^\circ_{\bar w} a_w +  g^2 ([X^\circ, \bar x]
+ [{\bar X}^\circ, x])+ 2 g^2 [ X^{\circ I}, x^I]  =0\label{gf}\,,
\ee
where $D^\circ$ stands for the covariant derivative with respect to
$A^\circ$.  Next we introduce the Faddeev-Popov ghost and antighost
fields $c$ and $\bar c$. They will be in the antisymmetric
representation of O(N), and will therefore be expanded as $a_w$ and
$a_{\bar w}$.  Then we add to the action the gauge fixing term
\be
S_{gf}=-\frac{1}{4\pi g^2}\int d^2w~{\cal G}^2
\ee
and the corresponding Faddeev-Popov ghost term
\be
S_{ghost}= \frac{1}{2\pi g^2} \int d^2w ~\bar c 
\frac {\delta {\cal G}}{\delta c} c\,,
\ee
where $\delta$ represents the gauge transformation with parameter $c$.

At this point, to single out the strong coupling limit of the action,
we rescale the fields in appropriate manner. 
\be
\begin{array}[b]{rclrcl}
A_w &=& g a_w^{\mathfrak t} + a_w^{\mathfrak n}\,, \qquad&
A_{\bar w} &=&  g a_{\bar w}^{\mathfrak t} + a_{\bar w}^{\mathfrak n}\,,\\[1ex]
c&=& g c^{\mathfrak t} + \sqrt g  c^{\mathfrak n}\,, \quad&
\bar c&=& g \bar c^{\mathfrak t} + \frac {1}{\sqrt g} \bar c^{\mathfrak n}\,.
\end{array}
\label{resc1}
\ee
and
\be
\begin{array}[b]{rclrcl}
X^i &=& X^{i(b)}+  \frac{1}{g}x^{\mathfrak n}\,,\qquad&
\theta_s &=& \theta_s^{\mathfrak t} + \frac{1}{\sqrt g}\theta_s^{\mathfrak n}\\[1ex]
\theta_c&=&\frac{1}{\sqrt {g}} \theta_c^{\mathfrak t} +
\frac{1}{\sqrt{g}}\theta^{\mathfrak n}\,, \qquad&
\chi &=& \chi^{\mathfrak t} + \frac {1}{g}\chi^{\mathfrak n}\,.
\end{array}\label{resc2}
\ee
The reason why we scale out $\theta_c^{\mathfrak t}$,
and not $\chi_1$, has been explained in section 2.1. We will make a further
comment on this point in the Appendix.

After these rescalings the action becomes
\be
S=S_{sc}+Q_{\mathfrak n}+ \cdots\label{stcoupl} \,,
\ee
where ellipsis denote non-leading terms in the strong coupling limit, 
\be
S_{sc}=
\frac{1}{\pi} \int_{{\cal C}_0} d^2w \,\Big\{\!\!&\Tr&\!\!\left[
\d_w x^{I{\mathfrak t}} \d_{\bar w} x^{I{\mathfrak t}} 
+ \d_w x^{{\mathfrak t}} \d_{\bar w} \bar x^{{\mathfrak t}} 
 +i \theta^{\mathfrak t}_s \d_{\bar w} \theta^{\mathfrak t}_s 
\right.\0\\
&&\left. -\d_w a^{\mathfrak t}_{\bar w}
\d_{\bar w} a^{\mathfrak t}_w - 
\d_w \bar c^{{\mathfrak t}} \d_{\bar w} c^{{\mathfrak t}}
\right]+i\chi^{\mathfrak t}\d_w \chi^{\mathfrak t}\Big\} \label{scaction}
\ee
and $Q_{\mathfrak n}$ is the purely quadratic term in the
$\phi^{\mathfrak n}$ fluctuations.  They can be integrated over and
give exactly 1. Although at first sight everything looks
like~\cite{bbn2}, there are several subtle differences.  In fact the
derivation of~(\ref{scaction}) and the integration of $Q_{\mathfrak
n}$ are not as straightforward as in~\cite{bbn2}. In order not to
interrupt the exposition we have collected the relevant details in the
Appendix.

In conclusion, in the strong coupling limit ($g\to\infty$),
the action becomes
\be
S_{sc}=
\frac{1}{\pi} \int_{{\cal C}_0} d^2w \,\Big\{\!\!&\Tr&\!\! \left[
\d_w x^{I{\mathfrak d}} \d_{\bar w} x^{I{\mathfrak d}} 
+ \d_w x^{{\mathfrak d}} \d_{\bar w} \bar x^{{\mathfrak d}} 
 +i \theta^{\mathfrak d}_s \d_{\bar w} \theta^{\mathfrak d}_s 
\right.\0\\
&&\left. +\d_w a^{\mathfrak d}_{\bar w}
\d_{\bar w} a^{\mathfrak d}_w + 
\d_w \bar c^{{\mathfrak d}} \d_{\bar w} c^{{\mathfrak d}}
\right]+ i\chi^{\mathfrak t} \d_ w \chi^{\mathfrak t}\Big\}\,.
\label{scaction1}
\ee
The matrices which appear in this action are diagonal $n\times n$
matrices, while $\chi^{\mathfrak n}$ is an $n$-component vector. We
can therefore rewrite the action in terms of the component modes
$\phi^{\mathfrak d} = \phi_{(1)},\ldots, \phi_{(n)}$:
\be
S_{sc}=
\frac{1}{\pi} \int_{{\cal C}_0} d^2w \,&\sum_{p=1}^n& \Big[
\d_w x^i_{(p)} \d_{\bar w} x^i_{(p)}
+i (\theta_{s(p)} \d_{\bar w} \theta_{s(p)} 
+ \chi_{(p)} \d_ w \chi_{(p)})\0\\
&& +\d_w a_{\bar w(p)}
\d_{\bar w} a_{w(p)} + 
\d_w \bar c_{(p)} \d_{\bar w} c_{(p)}
\Big]\,.
\label{scaction2}
\ee 
At first sight this looks like a theory of free fields on ${\cal
C}_0$.  However, as pointed out in~\cite{bbn2}, this is not correct.
The fields $x^i_{(p)}, ...$ are not single-valued on the cylinder: by
going around a branch point some of the $x^i_{(p)}$ is mapped to some
adjacent one, and this is precisely the way a branch point describes a
string interaction. The right interpretation was given in~\cite{bbn2}:
every $n$-tuple of fields $x^i_{(p)}$ $(\theta_{s(p)}, \chi_{(p)},
a_{w(p)},c_{(p)})$ form a unique well defined field $\tilde x^i$
$(\tilde\theta_s,\tilde\chi,\tilde a_w, \tilde c)$ on the Riemann
surface $\Sigma$, which is the covering of the cylinder defined by the
background we have expanded about. At this point we have a well
defined action on all of $\Sigma$, which is smooth also in
correspondence with the branch points. Therefore we can remove without
harm the regulator introduced before.

Finally we can write the strong coupling action~(\ref{scaction2}) as follows
\be 
S_{sc}^\Sigma&=& S^\Sigma_{het} + S^\Sigma_{Maxwell},\label{h+M}\\[2ex]
S^\Sigma_{het}&=&\frac{1}{\pi} \int_{\Sigma} d^2z \left( 
\d_z \tilde x^i\d_{\bar z} \tilde x^i
+i (\tilde\theta_{s} \d_{\bar z} \tilde\theta_{s} 
+\tilde \chi \d_z \tilde\chi)\right)\label{het}\\[1ex]
S^\Sigma_{Maxwell}&=& \frac{1}{\pi} \int_{\Sigma} d^2z\left(
g^{z\bar z}
\d_z \tilde a_{\bar z}
\d_{\bar z} \tilde a_{z} + 
\d_z \tilde {\bar c} \d_{\bar z}\tilde c\right).
\label{Max}
\ee
In~(\ref{het}) a $\sqrt{\omega_z}$ (resp. $\sqrt{\omega_{\bar z}}$)
factor has been absorbed in $\tilde{\theta_s}$
(resp. $\tilde\chi$) which is a $(\frac{1}{2},0)$
(resp. $(0,\frac{1}{2})$) differential on $\Sigma$ and the metric in
the Maxwell term is $g_{z \bar{z}}= \omega_z\omega_{\bar z}$.

Summarizing, what we have obtained in this subsection is that the
strong coupling effective theory is given by the Green-Schwarz
heterotic string action on the branched covering worldsheet plus a
decoupled Maxwell theory on the same surface.

\subsection{String amplitudes}

The structure of the strong coupling theory is now parallel to the one
obtained for the Matrix String Theory in~\cite{bbn2} and we can
quickly draw our conclusions by simply taking the results
of~\cite{bbn2,bbnt} and applying them to HMST. The Riemann surface
$\Sigma$, generated by the instanton, is taken to represent a
heterotic string process. The surface has punctures (i.e. the points
of the covering that correspond to $z=0$ and $z=\infty$) which
represent the incoming and outgoing strings: the length of each
asymptotic string, i.e. the multiplicity of the corresponding branch
point plus one~\cite{bbnt}, is physically interpreted as its
light-cone momentum $p_+$. Naturally, in order to fully describe the
asymptotic string states, we have to introduce in the path integral
the corresponding vertex operators. Each vertex is constructed out of
the $\tilde x, \tilde \theta, \tilde \chi$ fields and, moreover,
specifies the string transverse momentum. Since the vertices do not
depend on the Maxwell field and ghosts, we can integrate them out (the
non-Cartan modes have been integrated out above). Since the action is
free, the integration produces a ratio of determinants, which turns
out to be a constant. However we have to take account of the zero
modes for the fields that have been rescaled. The rescaled fields on
$\Sigma$~(\ref{resc1}) are
\begin{equation}
\tilde a_z\,\to g\, \tilde a_z\, ,\qquad \tilde a_{\bar z}\,\to g\, 
\tilde a_{\bar z}\, ,\qquad
\tilde c\,\to g\,\tilde c\, ,\qquad
\tilde {\bar c}\,\to g\, \tilde{\bar c}
\, .\0\pagebreak[3]
\end{equation}
As a consequence the Maxwell partition function rescales too with a
factor depending on the zero modes. By counting the zero modes of the
Maxwell fields and ghosts we conclude,\footnote{Fermionic zero modes
are absent if $2h+p-2>0$} as in~\cite{bbn2}, that (thanks to the
Maxwell partition function) the string amplitude corresponding to
$\Sigma$ is proportional to $g^{2-2h-p}$, where $h$ and $p$ are
respectively the number of handles and punctures of $\Sigma$. Being
the string coupling $g_s= g^{-1}$, this multiplicative factor is
exactly what we need in order to correctly reproduce string
interaction theory.

Of course this is not the end of the story. In order to obtain the
complete amplitude we have to sum over all the instantons of the same
topological type $(h,p)$.  Let us call $V_1,\ldots , V_k$ the vertex
operators corresponding to $k$ incoming and outgoing strings, and
insert them into the path integral.  The genus $h$ amplitude (in the
strong coupling limit) will schematically be:
\be
\langle V_1,\ldots,V_k\rangle_h= g_s^{-\chi}\int_{{\cal M}^{(h,n)}_{N}}dm~ 
\int {\cal D}[\tilde{x},\tilde{\theta},\tilde{a},\tilde{c}]  
V_1\ldots V_k\e^{-S_{het}}\,,\label{vertampl}
\ee
Integrating over ${{\cal M}^{(h,p)}_{N}}$ means integrating over all
distinct instantons which underlie the given string process for fixed
$N$, that is to say with assigned incoming and outgoing strings and
string interactions. In ordinary string interaction theory ${{\cal
M}^{(h,p)}}$ is nothing but the moduli space of Riemann surfaces of
genus $h$ with $p$ punctures, a complex space of dimension
$3h-3+p$. In HMST ${{\cal M}^{(h,p)}_{N}}$ is the same as in MST and
we can simply summarize the result obtained in~\cite{bbnt}: ${{\cal
M}^{(h,p)}_{N}}$ is a discrete slicing of the moduli space of Riemann
surfaces of genus $h$ with $p$ punctures, every slice having complex
dimension $2h-3+p$; in other words $h$ moduli are discrete in HMST; we
have argued in~\cite{bbnt} that when $N\to \infty$ these discrete
parameters become continuous and the full moduli space of Riemann
surfaces is recovered.

Therefore we can say that for large $N$ the strong coupling limit of
HMST reproduces the heterotic string interaction theory.  

As pointed out in section~\ref{sec:2}, in the absence of Wilson lines,
this is the $SO(32)$ theory compactified to nine dimensions on a
circle of very small radius.

To obtain other heterotic string theories, one must introduce Wilson
lines $\chi B\chi$, which at strong coupling become $\chi B\chi\to
\chi_0 B\chi_0$. The latter term is lifted to the covering as $\tilde
\chi B \tilde \chi$ and accounts for the breaking of $SO(32)$ to some
suitable subgroup.  As remarked at the beginning, with a standard
choice of Wilson lines, one can break $SO(32)$ to $SO(16)\times
SO(16)$, and relate the model to the {\it ten-dimensional} $E_8 \times
E_8$ string also broken to $SO(16)\times SO(16)$.

\appendix

\section{Appendix}

In this Appendix we discuss the derivation of the strong coupling
action~(\ref{stcoupl}) and the integration of the $Q_{\mathfrak n}$
therein. The problems we have to face are related to the presence in
the action of chiral fermions and to the absence in the HMST of half
the supersymmetry, compared to MST. In~(\ref{eSYM}) all fermions appear
quadratically, therefore they can be (at least formally) integrated
and give an overall well defined fermion determinant, since chiral
anomalies from different multiplets cancel. If we were able to
explicitly expand it in powers of $1/g$, we would start from this well
defined expansion and then integrate also over the
bosons. Unfortunately we do not know how to do that. Therefore we can
only carry out the expansion directly in the action.  However, if we
do so, we break the abovementioned well-defined determinant to pieces
and, in particular, we have to treat each chiral species separately.
This, in turn, leads to the longstanding problem of representing
chiral fermions in field theory. Various solutions have been proposed
over the years: dynamical mechanisms based on an infinite number of
flavors~\cite{NN} or on a generalized Pauli-Villars
regularization~\cite{FS} (see the reviews \cite{Neu,Fuji}, and the
references therein); in two dimensions, in particular, the twistor
formalism has been extensively used~\cite{twistor}. Any of these
choices would require a long technical detour from the mainstream of
our paper.  Therefore we take a simpler course. We operate formally on
the action, by making sure, however, that our procedure leads at every
step to a well-defined result.\footnote{Our wording is probably too
poor in regard to fermion determinants: by well-defined we mean that a
determinant is a function rather than a section of some nontrivial
line bundle.}  We warn the reader that a complete confirmation of the
correctedness of this way of proceeding would come from the full
expansion of the path integral in powers of $1/g$. This is however
beyond our present ability.

We deal first with the action term $\chi D_w\chi$, which turns out to be the
most delicate. Let us use~(\ref{chi}),~(\ref{diagonal}) and rewrite it as
\be
S_\chi=\frac{i}{\pi}\int\chi D_w\chi&= &i\int (\chi^{\mathfrak t} + 
\chi^{\mathfrak n})\left( \d_w - 
a_w^{\mathfrak t} -a_w^{\mathfrak n}\right) 
(\chi^{\mathfrak t} + \chi^{\mathfrak n})\0\\
&&=\frac{i}{\pi}\int\left(\chi^{\mathfrak t}\d_w\chi^{\mathfrak t}-
 \chi^{\mathfrak t}a_w^{\mathfrak n}
\chi^{\mathfrak t}-\chi^{\mathfrak t}a_w^{\mathfrak t}\chi^{\mathfrak n}-
\chi^{\mathfrak n}a_w^{\mathfrak t}\chi^{\mathfrak t}+\cdots\right)\0\\
&&=\frac{i}{\pi}\int\left(\chi_0\d_w \chi_0 -\chi_0 {\mathfrak a}_w \chi_0+ 
2\chi_0 a_w^{\mathfrak d}\chi_1
+\cdots \right).\label{chidichi}
\ee
Dots represent non-leading terms in the strong coupling limit once
the rescalings~(\ref{resc1}), (\ref{resc2}) are applied, so for
simplicity we drop them right away. However it should be kept in mind
that the rescalings will become effective only later on.  The matrix
${\mathfrak a}_w$ is $n\times n$ antisymmetric. Its elements are given by
$$
({\mathfrak a}_w)_{ij} = (a_w^{\mathfrak n})_{2i,2j} 
+(a_w^{\mathfrak n})_{2i,2j+1}+(a_w^{\mathfrak n})_{2i+1,2j}
+(a_w^{\mathfrak n})_{2i+1,2j+1}\,.
$$

Looking at the last line of~(\ref{chidichi}) a difficulty is
immediately evident. The second term should simply not be there in the
strong coupling limit, the first term should end up in $S_{sc}$ and
the last term should account for the overall result of 1 in the
integration of $Q_{\mathfrak n}$.  But first and last term are
intertwined. We have to disentangle them.  In so doing we will solve
also the problem of the second term.

To this end let us, for simplicity, single out the Grassmann path
integral involving the terms~(\ref{chidichi}).
$$
\int {\cal D}\chi_0{\cal D}\chi_1 ~e^{-S_\chi}\,.
\vspace*{.5ex}
$$
In this path integral we introduce a delta function
$\delta(\chi_0-\psi)$ where $\psi$ is a fermionic field of the same
type as $\chi_0$ (i.e. a vector with $n$ components, each of which is
a real spinor in the fundamental of $SO(32)$), and integrate over
$\psi$. Then we lift the delta function to the exponent by means of a
Lagrange multiplier $\l$, which is itself a spinor of the same type as
$\chi_0$ and $\psi$. The path integral is equivalent to the initial
one, provided we integrate over the modes of $\chi_0,\chi_1,\psi$ and
$\lambda$ and provided the action~(\ref{chidichi}) is replaced by
$$
S'_\chi=\frac{i}{\pi}\int(\chi_0\d_w \chi_0 -\psi {\mathfrak a}_w
\chi_0+2\psi a_w^{\mathfrak d}\chi_1 -\lambda \chi_0 +\lambda
\psi+\cdots)\,.
$$
Now we redefine $\chi_1$ as
$$
\chi_1'=\chi_1 -\frac{1}{2 a_w^{\mathfrak d}}\lambda - \frac{1}{2
a_w^{\mathfrak d}}{\mathfrak a}_w \chi_0
$$
and obtain
\be
S'_\chi = \frac{i}{\pi} \int (\chi_0\d_w \chi_0 + 2\psi a_w^{\mathfrak d}\chi'_1-\l \chi_0
+\cdots)\,.
\ee
Another way to get rid of the term $\chi_0 {\mathfrak a}_w \chi_0$ is
to integrate first over the $a_w^{\mathfrak n}$ degrees of freedom and
their conjugate: a well-known procedure to englobe linear terms in a
gaussian integration leads to a quartic term in $\chi_0$, which is
however vanishing.

Summarizing, the path integral involving the $\chi$ modes has become
\be
\int {\cal D}\chi_0{\cal D}\chi_1 {\cal D}\l {\cal D}\psi ~
e^{\frac{i}{\pi} \int (\chi_0\d_w \chi_0 + 
2\psi a_w^{\mathfrak d}\chi_1-\l \chi_0+\cdots)}\,.
\ee
It remains for us to rescale the fields according according to
(\ref{resc1}),~(\ref{resc2}) and, in addition, $\l$ according to
$\lambda\to \frac{1}{g}\l$. As a consequence, all terms represented by
dots, as well as the $\l\chi_0$ term, will drop out in the strong
coupling limit, leaving a zero volume integration constant. We will
comment about this later on. Disregarding it for a moment, the $\chi$
path integral in the $g\to \infty$ limit becomes
\be
\int {\cal D}\chi_0{\cal D}\chi_1{\cal D}\psi~e^{\frac{i}{\pi} 
\int (\chi_0\d_w \chi_0 - 2\psi a_w^{\mathfrak d}\chi_1)}\,.
\ee
Now the first term is shuffled to $S_{sc}$ in~(\ref{scaction}), while
the second term is exactly what we need in order to get 1 from the
path integration of $Q_{\mathfrak n}$. Let us concentrate now on the
latter.

$Q_{\mathfrak n}$ has the form
\be
Q_{\mathfrak n}&=& Q_{\mathfrak n}^{(matter)}+ Q_{\mathfrak n}^{(gauge)}\0\\[1ex]
Q_{\mathfrak n}^{(matter)}&=& \frac {1}{\pi} \int d^2w \Tr \left[ \bar 
x^{\mathfrak n}{\cal Q} x^{\mathfrak n}+  
x^{I{\mathfrak n}}{\cal Q} x^{I{\mathfrak n}}+ 
i (\theta_s^{\mathfrak n}, \theta_c^{\mathfrak n}) {\cal A}
\left(\matrix{\theta_s^{\mathfrak n}\cr \theta_c^{\mathfrak n}\cr}\right) 
- 2 \psi a_w^{\mathfrak d}\chi_1
\right]\label{Qn}\\
Q_{\mathfrak n}^{(gauge)}&=&-\frac {1}{\pi} \int d^2w \Tr \left[ 
a_{\bar w}^{\mathfrak n}{\cal Q} a_w^{\mathfrak n}+
\bar c^{\mathfrak n}{\cal Q} c^{\mathfrak n}\right],\0
\ee
where 
\be
{\cal Q} = {\rm ad}_{{\bar X}^\circ}\cdot {\rm ad}_{ X^\circ}-
 {\rm ad}_{a_{\bar w}^{\mathfrak t}}\cdot {\rm ad}_{a_w^{\mathfrak t}} +
 {\rm ad}_{x^{I{\mathfrak t}}}\cdot {\rm ad}_{x^{I{\mathfrak t}}}\0
\ee
and
\be
{\cal A} =\left( \matrix { - {\rm ad}_{a_{\bar w}^{\mathfrak t}} &i \gamma_i 
{\rm ad}_{X^{\circ i}} \cr
i\tilde \gamma_i {\rm ad}_{ {\bar X}^{\circ i}} & 
 {\rm ad}_{a_{ w}^{\mathfrak t}}\cr}\right)\0\,.
\ee
It is understood that one must integrate $Q_{\mathfrak n}$ over all the 
${\mathfrak n}$ modes, including $\chi_1$ (remember~(\ref{chi})), {\it and in 
addition} over $\psi$. In a very obvious way $\psi$ can be written as
$\psi^{\mathfrak t}= \frac {\psi}{\sqrt 2} \left(\matrix{1\cr 1\cr}\right)$
so that
$$
 2 \psi a_w^{\mathfrak d}\chi_1=- \psi^{\mathfrak t} a_w^{\mathfrak t}
\chi^{\mathfrak n}-\chi^{\mathfrak n} a_w^{\mathfrak t}
\psi^{\mathfrak t}\,.
$$

Our aim is to show that the integration over the bosonic degrees of
freedom in the path integral of $Q_{\mathfrak n}$ is exactly
compensated for by the integration over the fermionic ones. As for
$Q_{\mathfrak n}^{(gauge)}$ the argument is the same as in~\cite{bbn2}
and will not be repeated here: the integral over the gauge and ghost
modes gives 1.

As for the matter part, integrating formally we obtain a ratio of
determinants of ${\cal A}$ and ${\cal Q}$. However there are here some
subtleties that we have to explain.  In $Q_{\mathfrak n}^{(matter)}$
the operator ${\cal A}$ is a chiral operator: there is an asymmetry
between $\theta_s^{\mathfrak n}$ and $\theta_c^{\mathfrak n}$.  The
asymmetry is measured in the same way we do for chiral anomalies: we
compute the trace of the square generators in the symmetric and
adjoint representations of the Lie algebra ${\mathfrak o}(N)$. Setting
=1 the corresponding quantity in the fundamental representation, we
get $N+2$ in the symmetric representation and $N-2$ in the adjoint.
Therefore the determinant that comes from the formal integration over
these modes is not well defined.
 
To see this point more clearly we need a further splitting of the
non-Cartan degrees of freedom.
\begin{equation}
\phi^{\mathfrak n}= \phi^{\mathfrak d}_\tau\otimes \tau + 
\phi^{\mathfrak d}_\sigma\otimes \sigma + 
\phi^{\mathfrak n'}\,,
\label{splitting}
\end{equation}
where $\phi^{\mathfrak d}_\tau$ and $\phi^{\mathfrak d}_\sigma$ are diagonal 
$n\times n$ matrices,
$$ 
\tau =\left(\matrix{1 &0\cr 0 & -1\cr}\right), \qquad
\sigma= \left(\matrix{0&1\cr 1&0\cr}\right)
$$
and $\phi^{\mathfrak n'}$ denotes the remaining non-Cartan
directions.  It is clear that the components $\phi^{\mathfrak d}_\tau$
and $\phi^{\mathfrak d}_\sigma$ are present only in the fields $X^i$
and $\theta_s$, not in the others.

Now, computing the trace of the square generators in the symmetric and
adjoint representations, it is easy to see that the chiral asymmetry
of $(N+2)-(N-2)=4$ between the two representations is accounted for
exactly by the modes $\vartheta_\sigma^{\mathfrak d}$ and
$\vartheta_\tau^{\mathfrak d}$.  But, calculating the trace in the
fundamental representation, we see that this asymmetry is exactly
compensated for by the $\chi_1$ and $\psi$ modes.  (Naturally these
compensations are the same that cooperate to cancel the chiral anomaly
in the initial action~(\ref{eSYM})).
 
Therefore the operator ${\cal A}$, as it appears in~(\ref{Qn}) is
non-chiral and the corresponding determinant well defined. We can
therefore proceed from now on as in~\cite{bbn2}.  What we have to
evaluate is the ratio $\left((\det {\cal A})^{16}/(\det {\cal
Q})^8\right)^{2N^2+N}$. The involved operators do not have zero modes.
The ${\det {\cal A}}$ in the numerator should be understood as $\sqrt
{\det (-{\cal A} {\cal A}^\dagger)}$. But ${\cal A}{\cal A}^\dagger=
{\cal A}^\dagger {\cal A}= - {\cal Q}$. Therefore the net result of
integrating over the non-Cartan modes is 1. This is the result
expected from supersymmetry in the absence of zero modes.

As for the overall Jacobian arising from the rescalings
(\ref{resc1}),~(\ref{resc2}) and the one of $\l$, it gives rise to a
factor $g^{60n}$. This factor in the $g\to\infty$ limit is an infinite
factor that must be related to the abovementioned zero volume factor
due to integration over $\l$. Such infinite and zero factors do not
exist for finite $g$. They are an artifact of the $g\to\infty$
limit. Therefore it is sensible to assume that they just compensate
for each other and give a constant which we can choose to be 1.

Resuming now the considerations at the beginning of this section, we
remark that formal manipulations of the path integral have led us,
nevertheless, to a blameless definition of the integration over the
non-Cartan modes.  As for the surviving Cartan modes, lifted to the
covering they give rise to a well-defined theory, the heterotic
superstring.

\pagebreak[3]

Finally we would like to make a comment on the possibility, mentioned
in section 2.1, of an alternative strong coupling limit which would be
obtained by scaling away all the $\chi$ degrees of freedom while
keeping $\theta_c^{\mathfrak t}$. This would surprisingly lead to a
type IIA theory.  We have already ruled out above this possibility on
the basis of some some simple arguments. Here we would like to
strengthen the argument that the strong coupling limit should be a
chiral theory. This can be done by a request of anomaly matching
condition similar to 't Hooft's one~\cite{thooft}.  The `flavor'
symmetry of our theory at $g=0$ can be thought of as a global
$SO(8)_R\times SO(8)_L\times SO(32)_L$. Suppose we gauge this symmetry (for
instance, by coupling $\theta_s$ to a gauge potential $C_w^{ij}$ via
$\theta_s C_w^{ij}\gamma_{ij}\theta_s$, and so on); it is elementary
to see that the corresponding anomalies are reproduced in the
heterotic strong coupling limit with the same relative coefficient,
while they would not reproduced in the would-be IIA strong coupling
limit.

\acknowledgments 

We would like to acknowledge valuable discussions we had with
G. Ferretti, E. Gava, A.~Hammou, F. Morales. We would also like to
thank L. Motl for interesting comments on the first version.  This
research was partially supported by EC TMR Programme, grant
FMRX-CT96-0012, and by the Italian MURST for the program ``Fisica
Teorica delle Interazioni Fondamentali''.

After completing this work, we saw a new paper by
T. Wynter~\cite{W_new} dealing with the strong coupling of MST.


\begin{thebibliography}{99}

\bibitem{DF} 
U.H. Danielsson and G. Ferretti, {\it The heterotic life of the
D-particle}, \ijmpa{12}{1997}{4581} [\hepth{9610082}].

\bibitem{KS}
S. Kachru and E. Silverstein, {\it On gauge bosons in the Matrix model approach 
to M theory}, \plb{396}{1997}{70} [\hepth{9612162}].
\bibitem{lowe}
D.A. Lowe, {\it Bound states of type I' D-particles and enhanced gauge symmetry},
\npb{501}{1997}{134} [\hepth{9702006}].

\bibitem{BM} 
T. Banks and L. Motl {\it Heterotic strings from
matrices}, \jhep{12}{1997}{004} [\hepth{9703218}].

\bibitem{rey}
S.-J. Rey, {\it Heterotic M(atrix) strings and their interactions},
\npb{502}{1997}{170} [\hepth{9704158}].

\bibitem{horava}
P. Horava, {\it Matrix theory and heterotic strings on tori}, 
\npb{505}{1997}{84} [\hepth{9705055}].

\bibitem{kabat}
D. Kabat and S.-J. Rey, {\it Wilson lines and T-duality in heterotic M(atrix) 
theory}, \npb{508}{1997}{535} [\hepth{9707099}.]

\bibitem{govin}
S. Govindarajan, {\it Heterotic M(atrix) theory at generic points in Narain 
moduli space}, \npb{507}{1997}{589} [\hepth{9707164}].

\bibitem{krogh}
M. Krogh, {\it A Matrix model for heterotic Spin(32)/$Z_2$ and type I string 
theory}, \npb{541}{1999}{87} [\hepth{9801034}].

M. Krogh, {\it Heterotic matrix theory with Wilson lines on the lightlike circle}
\npb{541}{1999}{98} [\hepth{9803088}].

\bibitem{BGS}
C.P. Bachas, M.B. Green and A. Schwimmer, {\it (8,0) Quantum mechanics and 
symmetry enhancement in type I' superstrings}, \jhep{01}{1998}{006} [\hepth{9712086}].

\bibitem{wynter}
T. Wynter, {\it Gauge fields and interactions in matrix string theory},
\plb{15}{1997}{349} [\hepth{9709029}].

\bibitem{GHV}
S.B.~Giddings, F.~Hacquebord, H.~Verlinde,
{\it High Energy Scattering of D-pair Creation in Matrix String Theory},
\npb{537}{1999}{260} [\hepth{9804121}].

\bibitem{bbn1}
G.~Bonelli, L.~Bonora and F.~Nesti, {\it Matrix string theory,
$2D$ instantons and affine Toda field theory}, \plb{435}{1998}{303} 
[\hepth{9805071}].

\bibitem{bbn2}
G. Bonelli, L. Bonora and F. Nesti,
{\it String Interactions from Matrix String Theory},
\npb{538}{1999}{100}  [\hepth{9807232}].

\bibitem{bbnt}
G. Bonelli, L. Bonora, F. Nesti and A. Tomasiello,
{\it Matrix String Theory and its Moduli space}, to appear in {\it
Nucl.\ Phys.}\ {\bf B} [\hepth{9901093}]

\bibitem{winter2} T. Wynter, {\it Anomalies and large N limit in matrix string 
theory}, \plb{439}{1998}{37} [\hepth{9806173}].

\bibitem{BCDP} M. Bill\`o, M. Caselle, A. D'Adda and P. Provero, {\it Matrix string states
in pure 2d Yang Mills theories}, \npb{543}{1999}{141} [\hepth{9809095}].

\bibitem{kostov} I.K. Kostov and P. Vanhove, {\it Matrix string partition 
functions}, \plb{44}{1998}{196} [hepth{9809130}].

\bibitem{makee} Yu.M. Makeenko, {\it Formulation of matrix theory at finite
temperature}, \hepth{9903030}.

\bibitem{gri} G. Grignani and G.W. Semenoff {\it Thermodynamic Partition Function
of Matrix Superstrings}, \hepth{9903246}.

\bibitem{sugino} F. Sugino, {\it Cohomological Field Theory Approach to Matrix 
Strings}, \hepth{9904122}.
 
\bibitem{motl}
L.~Motl, {\it Proposals on Nonperturbative Superstring Interactions},
\hepth{9701025}.

\bibitem{BS}
T.~Banks and N.~Seiberg,
{\it Strings from Matrices}, \npb{497}{1997}{41} [\hepth{9702187}].

 
\bibitem{DVV}
R.~Dijkgraaf,  E.~Verlinde, H.~Verlinde,
{\it Matrix String Theory},
\npb{500}{1997}{43} [\hepth{9703030}].

\bibitem{HV}
H.~Verlinde, {\it A Matrix String Interpretation of the Large N Loop Equation},
\hepth{9705029}.

\bibitem{BC}
L.~Bonora, C.S.~Chu,
{\it On the String Interpretation of M(atrix) Theory}, 
\plb{410}{1997}{142} [\hepth{9705137}].

\bibitem{GSW} 
M.B.~Green, J.H.~Schwarz, E.~Witten,
{\it Superstring Theory}, Cambridge Univ.\ Press, Cambridge 1987, vol.\ I.

\bibitem{PW} 
J. Polchinski and E. Witten, {\it Evidence for heterotic--type I
string duality}, \npb{460}{1996}{525} [\hepth{9510169}].

\bibitem{HW} 
P. Horava amd E. Witten, {\it Heterotic and type I string dynamics
from eleven dimensions}, \npb{460}{1996}{506} [\hepth{9510209}].

\bibitem{NN} R. Narayanan and H. Neugberger, {\it Infinitely many regulator 
fields for chiral fermions}, \plb{302}{1993}{62} [\heplat{9212019}].

\bibitem{FS} S.A. Frolov and A.A. Slavnov, \plb{309}{1993}{344}.

\bibitem{Neu} H. Neuberger, {\it A lecture on chiral fermions}, \heplat{9511001};

\bibitem{Fuji} 
K. Fujikawa, {\it Analytic index and chiral fermions}, {\it Indian
J. Phys.} {\bf 70A} (1996) 275-291 [\hepth{9506003}].

\bibitem{twistor} 
M. Tonin, {\it $\kappa$-symmetry as world sheet supersymmetry of D=10
heterotic superstring}, \ijmpa{t}{1992}{6013};

F. Delduc, A. Galperin, P. Howe and E. Sokatchev, {\it A twistor formulation
of the heterotic D=10 superstring with manifest (8,0) world sheet
supersymmetry}, \prd{47}{1992}{578} [\hepth{9207050}];

S. Aoyama, P. Pasti and M. Tonin, {\it The GS and NRS heterotic strings from 
twistor string models}. \plb{283}{1992}{213};

D. Sorokin and M. Tonin, {\it On the chiral fermions in the twistor-like 
formulation of D=10 heterotic string}, \plb{326}{1994}{84} [\hepth{9212019}];

P. Howe, {\it A note on chiral fermions and heterotic strings}, 
\plb{332}{1994}{61} [\hepth{9403177}];

E. Ivanov and E. Sokatchev, {\it Chiral fermion action with (8,0) worldsheet
supersymmetry}, \hepth{9406071}.

\bibitem{thooft} G. 't Hooft, {\it Naturalness, chiral symmetry and spontaneous
chiral symmetry breaking} in {\it Dynamical Symmetry Breaking}, edited by 
A. Fahri and R. Jackiw, World Scientific, 1982.

\bibitem{W_new} 
T. Wynter, \emph{High energy scattering amplitudes from Matrix string
theory}, \hepth{9905087}.

\end{thebibliography}
\end{document}